\documentclass[pra,twocolumn,tightenlines,showpacs,nofootinbib]{revtex4}
\usepackage{bm,dcolumn,amsmath,graphicx,amsfonts,amssymb}

\newcommand{\eref}[1]{Eq.~(\ref{#1})}
\newcommand{\tref}[1]{Table~\ref{#1}}

\begin{document}
%####################################################################
\title{Study of the correlation effects in Yb$^+$ and implications for
parity violation}
\author{S.~G.~Porsev$^{1,2}$}
\author{M.~S.~Safronova$^1$}
\author{M.~G.~Kozlov$^{2,3}$}
\affiliation{
$^1$Department of Physics and Astronomy, University of Delaware,
    Newark, Delaware 19716, USA\\
$^2$Petersburg Nuclear Physics Institute, Gatchina,
    Leningrad District, 188300, Russia\\
$^3$St. Petersburg Electrotechnical University ``LETI'', Prof. Popov Str. 5,
    St. Petersburg, 197376, Russia}

\date{ \today }

\begin{abstract}
Calculation of the energies, magnetic dipole hyperfine structure constants, E1 transition
amplitudes between the low-lying states, and nuclear
spin-dependent parity-nonconserving amplitudes for the $^2\!S_{1/2} - \, ^2\!D_{3/2,5/2}$
transitions in $^{171}$Yb$^+$ ion is performed using two
different approaches. First, we carried out many-body perturbation
theory calculation considering  Yb$^+$  as a monovalent system.
Additional all-order calculations are carried out for selected
properties. Second, we carried out configuration interaction
calculation considering Yb as a 15-electron system and  compared the
results obtained by two methods. The accuracy of different methods is
evaluated. We find that the monovalent description is inadequate for
evaluation of some atomic properties due to significant mixing of
the one-particle and the hole-two-particle configurations. Performing
the calculation by such different approaches allowed us to establish
the importance of various correlation effects for Yb$^+$ atomic
properties for future improvement of theoretical precision in this
complicated system.
\end{abstract}
\maketitle

%####################################################################
\section{Introduction}
%####################################################################
The Yb$^+$ ions have been a subject of heightened  interest in recent
years owing to use of this system in a number of different
applications including quantum information
studies~\cite{OlmYouMoe07,WanJohFen11}, searches for variations of
fundamental constants~\cite{PeiLipSch04}, and development of the
optical frequency
standards~\cite{SchDanNgu09,TamWeyLip09,MeySteRat12,HunOkhLip12}.

Manipulation and detection of a trapped  Yb$^{+}$  hyperfine qubit
was described in~\cite{OlmYouMoe07}. An efficient scheme to carry out
gate operations on an array of trapped Yb$^{+}$ ions was suggested
in~\cite{WanJohFen11}. Yb$^+$ is of particular interest to the atomic
clock development  due to the availability of two different
(quadrupole \cite{TamWeyLip09} and octupole \cite{HunOkhLip12})
metastable transitions that can be used as optical frequency
standards. In 2012, the performance of the optical  frequency
standard based on electric-octupole transition $^2\!S_{1/2}(F=0)
\rightarrow \, ^2\!F^o_{7/2}(F=3)$ in a single trapped Yb$^+$ ion was
investigated  \cite{HunOkhLip12}. This work has demonstrated that the
octupole transition in $^{171}$Yb$^+$ can be used to realize an
optical clock with a systematic uncertainty of $7.1\times 10^{-17}$
\cite{HunOkhLip12}. Moreover, it has been shown that a clock based on
a linear combination of the quadrupole and the octupole transition
frequencies of Yb$^+$ can have a significantly reduced blackbody
shift \cite{YudTaiOkh11}. An availability of two metastable
transitions suitable for the development of the precision frequency
standard made Yb$^+$ an attractive candidate for the search of the
variation of the fine-structure constant.

The $^2\!S_{1/2} \rightarrow \, ^2\!D_{3/2}$ transition in Yb$^+$ was
also proposed~\cite{TorSchZha10} for study of the nuclear
spin-dependent (SD) parity-nonconserving (PNC) effects. Such an
experiment will be able to yield the nuclear anapole moment that
arises due to parity-violating interaction between nucleons in the
nucleus ~\cite{Wood97}. Study of the weak hadronic interactions is of
particular interest due to significant discrepancy between
constraints on weak nucleon-nucleon couplings obtained from the
cesium anapole moment and those obtained from other nuclear
parity-violating measurements ~\cite{HaxWie01,SafPalJia01}.

Accurate calculation of Yb$^+$ properties is very difficult owing to
the large number of low-lying states of the hole-two-particle
configurations such as $4f^{13}5d6s$ and their strong mixing with
one-particle (monovalent) configurations, such as $4f^{14}6p$. The
properties of ytterbium ions were studied in a number of theory
papers (see, e.g.,~\cite{SafSaf09} and references therein). Because
the main configuration of the ground state of Yb$^+$ is $4f^{14} 6s$,
this ion can be considered as a system with one electron above the
closed shells. Alternatively, one can treat the $4f$ electrons as the
valence electrons, and consider Yb$^+$ as a system with 15 valence
electrons. Both approaches have advantages and drawbacks. In the following
we refer to them as a single-electron approach and a many-electron approach.

The advantages of a monovalent (single-electron) method are high accuracy and relative simplicity.
In particular, the core-valence correlations can be treated very
accurately. However, the problem is that the states belonging to the configurations with
the unfilled $4f$ shell, such as $4f^{13} 6s^2$ and $4f^{13} 5d6s$,
are lying very low in Yb$^+$. A knowledge of their properties is
important for a number of experimental schemes mentioned above. A
single-electron method is unable to treat such states since these are
not monovalent states. Moreover, the states with filled $4f$ shell
(like $4f^{14}6p \,\, ^2\!P^o_{3/2}$) can strongly interact with a
closely located state with the unfilled $4f$ shell. This mixing can
significantly affect the properties of both states. Again, the
single-electron approach does not take into account this interaction
that drastically affects the accuracy of this approach for the states
where this mixing is large. This effect is illustrated on the example
of calculation of the magnetic dipole hyperfine structure (hfs)
constant $A$ of the $4f^{14}6p \,\, ^2\!P^o_{3/2}$ state. It was
calculated by several different methods that considered Yb$^+$ as a
monovalent system~\cite{MarGouHan94,SafSaf09,DzuFla11,SD11}.

The resulting values are in reasonable agreement with each other but are
factor of two smaller than the experimental result. As we show in
this work, the reason for this discrepancy of theory and experiment
is the strong configuration interaction between the $4f^{14}6p \,\,
^2\!P^o_{3/2}$ and $4f^{13}5d6s \,\, ^3[3/2]^o_{3/2}$ states.

The many-electron methods, such as the conventional configuration
interaction (CI), allow us to study  the properties of the states
with both filled and unfilled $4f$ shell on equal footing. It also
allows to take into account the configuration interaction between nearby levels.
However, the accuracy of the 15-electron CI approach is generally
lower than that of single-electron methods due to omission of the
correlation corrections between the core [$1s^2,...,5p^6$] electrons and
the valence electrons. So far, it has not been possible to incorporate
successfully the core-valence correlations into a many-electron CI.

In this work, we carried out calculation of Yb$^+$ properties using
both the single-electron approach, with both many-body perturbation
theory (MBPT) and all-order methods, and the 15-electron
configuration interaction method. The use of the both approaches
allows us to study the properties of {\it all} low-lying states.
Since these methods are to some extent complementary to each other,
they give us  clearer picture of the importance of various
correlation effects and provide more complete theoretical description
of the Yb$^+$ properties. This work will allow to outline a pathway
for the development of more accurate approaches for the calculation
of  Yb$^+$ properties of interest to applications listed above.
Because of the importance of the Yb$^+$ for various applications,
experimental values  of other properties will become available in the
future for further theory tests. Yb$^+$ is an excellent system for
benchmark tests of new theoretical approaches capable of describing
strong electron correlations. A development of such new approaches is
also needed to improve theoretical description of complex atoms, such
as Dy or Ho, which is becoming more important owing to recent
experimental developments and new proposals with these
systems~\cite{CinLeeFer08,SafMol08,LuBurYou11}.

Another goal of this paper is to evaluate spin-dependent
parity-violating amplitudes for the $4f^{14}6s \,\, ^2\!S_{1/2} -
4f^{14}5d \,\, ^2\!D_{J}$ transitions in Yb$^+$ and to  study the
effects of various correlation corrections to this quantity. The
calculation of the PNC amplitude is required to analyze the
experimental PNC studies and extract the anapole moment (unless the
measurements are carried out with several isotopes). Such
experimental study with Yb$^+$ is presently underway in Los Alamos
~\cite{TorSchZha10}. So far, a non-zero anapole moment was observed
only in Cs~\cite{Wood97}. The Cs result is in disagreement with the
nuclear physics predictions for the Cs anapole moment and constraints
on weak nucleon-nucleon couplings~\cite{HaxWie01,SafPalJia01}
prompting further investigations. The spin-dependent PNC effects in
the $4f^{14}5d \,\, ^2\!D_{3/2} - 4f^{14}6s \,\, ^2\!S_{1/2}$
transition of Yb$^+$ were recently investigated in~\cite{DzuFla11}.
The authors of \cite{DzuFla11} treated Yb$^+$ as a monovalent system.
They noted significant cancelation between different terms
contributing to the SD PNC amplitudes which merited further
investigation carried out here. We note that the total uncertainty in
the value of the anapole moment that can be extracted from the
experiment with a single isotope includes the theoretical and
experimental uncertainties.

Other PNC experiments that are presently underway include studies
with Yb~\cite{TsiDouFam09}, Fr~\cite{SheOroGom10}, and
Ra$^+$~\cite{VerGirWan10}. Large atomic parity violation effect was
observed in neutral Yb ~\cite{TsiDouFam09}.

The paper is organized as follows. The Sec.~\ref{single} is devoted to
the single-electron approach. We present the results of calculations of the
low-lying energy levels, the magnetic dipole hfs constants, E1 transition
amplitudes between the low-lying states, and nuclear spin-dependent
parity-nonconserving amplitudes for the $^2\!S_{1/2} - \, ^2\!D_{3/2,5/2}$
transitions. The all-order results are also given for the energy levels
and electric dipole matrix elements. In Sec.~\ref{CI} we present
the energy levels, magnetic dipole hfs constants, and E1
transition amplitudes for the low-lying states calculated in the
framework of the 15-electron CI method. We compare the results obtained
by either method. If the results differ from each other the reasons
are analyzed. The conclusions are described in Sec.~\ref{Concl}. We
use atomic units $\hbar = |e| = m_e = 1$ thorough the paper unless
stated otherwise.
%==============================================
\section{Single-electron approach and results}
\label{single}
%==============================================

In this approach, the $4f$ electrons are considered as the core
electrons. We start from the solution of the Dirac-Fock (DF)
equations carrying out the self-consistency procedure for the
[1$s^2$,...,4$f^{14}$] core electrons. Then, the valence orbitals
$6-8s$, $6-7p$, and $5-6d$ were constructed in the V$^{N-1}$
approximation ($N$ is the total number of the electrons in the
system). The basis set used in calculations included virtual orbitals
up to $23s, 23p, 23d, 22f$, and $14g$ formed with the help of the
recurrent procedure described in Refs.~\cite{Bog91,KozPorFla96}. The
MBPT corrections can be included by solving the equation
\begin{equation}
H_{\rm eff}\, \psi_n = \varepsilon_n  \psi_n
\end{equation}
with the effective Hamiltonian defined as $H_{\rm eff} \equiv H_0 +
\Sigma$, where $H_0$ is the Dirac-Fock Hamiltonian and the operator
$\Sigma$ takes into account virtual core excitations.
\subsection{Energy levels}
First, we find the energies of the low-lying states in various
approximations and compare them with the experimental values~\cite{NIST}.
As we already mentioned, we are able to obtain only the energies for
the states with filled $4f$ shell in the framework of this approach.
In~\tref{tab_E6s} we present the ionization potential (line 1) and the
energies of the low-lying states obtained in different approximations.
At the DF stage of the calculations even the order of the levels is
incorrect. An inclusion of the second-order MBPT corrections
restores the correct order of levels listed in Table~\ref{tab_E6s}.
The second-order MBPT values are listed in column labeled ``MBPT''.
%###################################################################
\begin{table}
\caption{The comparison of the energy levels calculated in different
approximations with experiment ~\cite{NIST}. The ionization potential
is given in the first line (in a.u.), the energy levels of the excited states
are counted from the ground state (in cm$^{-1}$). The columns labeled ``DF'' and
``MBPT'' correspond to the Dirac-Fock and DF+MBPT approximations with the
MBPT corrections included in the second order. The higher orders of
the MBPT are included in the results listed in the column labeled ``MBPT(HO)''.
The results of the single-double all-order calculation are listed in
the column labeled ``All-order''.}
\label{tab_E6s}
\begin{ruledtabular}
\begin{tabular}{lcccccc}
               &\multicolumn{1}{c}{DF}
                          &\multicolumn{1}{c}{MBPT}
                                     &\multicolumn{1}{c}{MBPT(HO)}
                                                &\multicolumn{1}{c}{All-order}
                                                         &\multicolumn{1}{c}{Experiment}  \\
\hline
$^2\!S_{1/2}$   & 0.41366 & 0.45211  & 0.44473  &0.45090 &0.44775\footnotemark[1] \\
$^2\!D_{3/2}$   &  24272  &  24450   &  22711   &22820   & 22961  \\
$^2\!D_{5/2}$   &  24752  &  25952   &  24178   &24261   & 24333  \\
$^2\!P^o_{1/2}$ &  24702  &  28636   &  27945   &27945   & 27062  \\
$^2\!P^o_{3/2}$ &  27513  &  32242   &  31403   &31481   & 30392  \\
\end{tabular}
\footnotemark[1]{This is equal to the ionization potential
 = 98269 cm$^{-1}$~\cite{NIST}.}
\end{ruledtabular}
\end{table}
%###################################################################

%The $\Sigma$ part is constructed using the second-order perturbation theory~\cite{DzuFlaKoz96b}.

As expected, the agreement between the experimental and theoretical
energies significantly improves with the inclusion of the correlation
corrections beyond DF approximation. At the same time, the
correlations are large and accounting for only the 2-nd order MBPT
corrections is not sufficient. To calculate the energy levels (and
subsequently other properties) more accurately, we need to take into
account the higher orders (HO) of the MBPT. We designate this
approximation as MBPT(HO) and label the results of such calculations
accordingly in the text and the tables.

In this approach, we include higher-order corrections by introducing
screening coefficients $C_k$ for the Coulomb lines in self-energy
diagrams (see, e.g.,~\cite{DzuKozPor98}). The latter can be
calculated as an average screening of the two-electron Coulomb radial
integrals of a given multipolarity $k$. These coefficients serve as
an approximation to the insertion of polarization operator in Coulomb
lines. The coefficients $C_k$ were chosen as follows: $C_0=1.3$,
$C_1=0.75$, and $C_k=1$ for $k \ge 2$. The resulting energies are
listed in Table~\ref{tab_E6s} in column labeled ``MBPT(HO)''. The
ionization potential obtained in the MBPT(HO) approximation agrees
with the experiment at the level of 0.07\%; the energies of the
even-parity states are within 1\% from the experimental values, and
the energies of the $^2\!P^o_{1/2}$ and $^2\!P^o_{3/2}$ states were
reproduced with the 3\% accuracy.

 As a test of the MBPT(HO) approach, we also calculated the energy levels
using the linearized single-double coupled-cluster method (also
referred to as the all-order method). This method allows to include
the higher-order correlation corrections in an \textit{ab initio} way
by effectively summing the dominant correlation contributions to
all orders of the perturbation theory. The single-double all-order method
was demonstrated to produce very accurate atomic properties for
alkali-metal atoms and other monovalent systems. We refer the reader
to the review ~\cite{SafJoh08} for a description of  the all-order
approach and its applications. The all-order data are listed in
column labeled ``All-order'' in Table~\ref{tab_E6s}. These energy values
have been previously listed in \cite{GhaEliSaf11}. We find that
\textit{ab initio} all-order energy levels are close to the
MBPT(HO) values serving as an additional verification of the MBPT(HO)
approximation.
% ---------------------------------------------------------------------
\subsection{Magnetic dipole hfs constants and E1 transition amplitudes}
% ---------------------------------------------------------------------
To calculate magnetic dipole hfs constants and E1 transition
amplitudes, we  construct effective valence operators for the
magnetic dipole hyperfine interaction $H_{\rm hfs}$ and the electric
dipole operator $d$. First, we solve the random-phase approximation
(RPA) equations, which is equivalent to the summation of the dominant
sequence of many-body diagrams to all orders of
MBPT~\cite{DzuKozPor98,KolJohSho82}. Then, we include additional
corrections (beyond RPA) to the effective operators:
the core-Brueckner, structural radiation (SR), and normalization corrections.

The results obtained for the hfs constants are listed in~\tref{HFS}.
%####################################################################
\begin{table*}
\caption{The breakdown of different contributions to the hfs
constants $A$ (in MHz) ($I=1/2$, $\mu$ = 0.4919~\cite{webel}). First
row gives the DF values and the following rows give MBPT(HO), RPA,
core-Brueckner ($\sigma$), structural radiation (SR), and
normalization (Norm.) corrections. The row labeled ``Total'' gives the final
numbers. The values are compared with the experimental and other
theoretical~\cite{SafSaf09,MarGouHan94,DzuFla11,SD11} results.}
\label{HFS}
\begin{ruledtabular}
\begin{tabular}{lcccccc}
&\multicolumn{1}{c}{$^2\!S_{1/2}$}
&\multicolumn{1}{c}{$^2\!D_{3/2}$}
&\multicolumn{1}{c}{$^2\!D_{5/2}$}
&\multicolumn{1}{c}{$^2\!P^o_{1/2}$}
&\multicolumn{1}{c}{$^2\!P^o_{3/2}$} \\
\hline
DF         &   9577  &  290  &  111  &  1542  &  183 \\
MBPT(HO)   &   2993  &  109  &   38  &   549  &   58 \\
RPA        &   1672  &  -55  & -308  &   323  &  132 \\
$\sigma$   &   -762  &  -13  &   -5  &   -10  &   -5 \\
SR         &   -188  &  167  &   67  &    -9  &  -35 \\
Norm.      &   -201  &   -9  &    1  &   -24  &   -3  \\
Total      &  13091  &  489  &  -96  &  2371  &  330  \\[0.1pc]
Experiment &  12645(2)\footnotemark[1]
                     &  430(43)\footnotemark[2]
                             & -63.6(7)\footnotemark[3]
                                     &  2104.9(1.3)\footnotemark[1]
                                              & 877(20)\footnotemark[4] \\[0.1pc]
Ref.~\cite{SafSaf09}
           & 13172   &       &       &  2350  & 311.5 \\
Ref.~\cite{MarGouHan94}
           & 12730   &       &       &  2317  & 391   \\
Ref.~\cite{DzuFla11}
           & 13217   &  291  &       &  2533  & 388   \\
Ref.~\cite{SD11}
           & 13332   &  447  &  -48  &  2516  & 322   \\
\end{tabular}
\footnotemark[1]{Reference~\cite{MarGouHan94}};
\footnotemark[2]{Reference~\cite{Ita06} and references therein};
\footnotemark[3]{Reference~\cite{RobTayGat99}};
\footnotemark[4]{Reference~\cite{BerMal92}}.
\end{ruledtabular}
\end{table*}
%####################################################################
This table illustrates that the MBPT corrections are generally
large and contribute significantly to the hfs constants $A$. The RPA,
core-Brueckner, SR, and normalization corrections also have to be
taken into account. The $^2\!D_J$ states are particular sensitive to
different corrections. For instance, the RPA correction even changes
the sign of $A(^2\!D_{5/2})$. The SR corrections (which are usually
relatively small) are found to be significant in this case and change
the values of $A(^2\!D_{3/2})$ and $A(^2\!D_{5/2})$ by more than
40\%.

The final values of all hfs constants obtained in this work are, in
general, in reasonable agreement with the experimental data and other
theoretical results with the exception of two cases. We find a
significant difference  between our value of $A(^2\!D_{3/2})$ and the
value found in~\cite{DzuFla11}. The difference is most probably due
to inclusion of the corrections beyond RPA in the present work. Our
result is in a good agreement with the experiment. All of the
theoretical values are in disagreement with the experimental value of
the $A(^2\!P^o_{3/2})$ demonstrating that this hfs constant cannot be
correctly reproduced in the framework of a single-electron approach.
As we will discuss in more detail in the section devoted to the
15-electron CI, this problem arises due to a strong mixing of the
$4f^{14}6p \,\, ^2\!P^o_{3/2}$ state and a nearby $4f^{14}5d6s \,\,
^3[3/2]^o_{3/2}$ state. A possible sensitivity of $A(^2\!P^o_{3/2})$
to the configuration mixing was also mentioned in
Ref.~\cite{DzuFla11}.

We also calculated the $E1$ amplitudes for the transitions between the low-lying states
and compared them with other available data. The lifetime $\tau$ of the $^2\!P^o_{1/2}$ state
 was measured
with a high precision in Ref.~\cite{OlmHayMat09}  to be equal to
8.12(2)~ns. The $^2\!P^o_{1/2}$ state can decay to the $^2\!D_{3/2}$
and $^2\!S_{1/2}$ states. The decay channel to the ground state
strongly dominates. The branching ratio from the $^2\!P^o_{1/2}$
state to the metastable $^2\!D_{3/2}$ state was measured to be
0.00501(15)~\cite{OlmYouMoe07}. Using two these quantities, we find
the transition probabilities
\begin{eqnarray}
W(^2\!P^o_{1/2} \rightarrow \, ^2\!S_{1/2}) &=& 0.995/\tau(^2\!P^o_{1/2}) \nonumber \\
                                            &=& 1.23(5) \times 10^8\, {\rm s}^{-1}
\end{eqnarray}
and
\begin{eqnarray}
W(^2\!P^o_{1/2} \rightarrow \, ^2\!D_{3/2}) &=& 0.00501/\tau(^2\!P^o_{1/2}) \nonumber \\
                                            &=& 6.17(18) \times 10^5\, {\rm s}^{-1}.
\end{eqnarray}
Respectively, the ``experimental'' reduced matrix elements (MEs) of
the electric-dipole moment operator are $|\langle ^2\!S_{1/2} ||d||
^2\!P^o_{1/2} \rangle| = 2.471(3)$ a.u. and $|\langle ^2\!D_{3/2}
||d|| ^2\!P^o_{1/2} \rangle| = 2.97(4)$ a.u.

At the present time, the most precise measurement of the
$^2\!P^o_{3/2}$ lifetime,  $\tau(^2\!P^o_{3/2}) = 6.15(9)$ ns, was
carried out in~\cite{PinBerJi94}. The $^2\!P^o_{3/2}$ state mainly
decays by the E1 transitions to the $^2\!S_{1/2}$, $^2\!D_{3/2}$, and
$^2\!D_{5/2}$ states. Then
\begin{eqnarray}
\frac{1}{\tau(^2\!P^o_{3/2})} &\approx& W(^2\!P^o_{3/2} \rightarrow \, ^2\!S_{1/2}) +
W(^2\!P^o_{3/2} \rightarrow \, ^2\!D_{3/2})  \nonumber \\
&+& W(^2\!P^o_{3/2} \rightarrow \, ^2\!D_{5/2}).
\label{W}
\end{eqnarray}

The transition probabilities $W(^2\!P^o_{3/2} \rightarrow \,
^2\!D_{3/2})$ and $W(^2\!P^o_{3/2} \rightarrow \, ^2\!D_{5/2})$ were
calculated in Ref.~\cite{SafSaf09} to be $3.6 \times 10^5$ s$^{-1}$
and $1.9 \times 10^5$ s$^{-1}$, respectively. Thus, they are more
than two orders of magnitude smaller than $1/\tau(^2\!P^o_{3/2})
\approx 1.626 \times 10^8$ s$^{-1}$. Even if the accuracy of
$W(^2\!P^o_{3/2} \rightarrow \, ^2\!D_J)$ is not so high (for
example, $\sim$ 20-30\%), it practically does not affect the final
accuracy of the $W(^2\!P^o_{3/2} \rightarrow \, ^2\!S_{1/2})$
inferred from the experiment. Using the experimental value of
$\tau(^2\!P^o_{3/2})$, we find  the probability of the $^2\!P^o_{3/2}
\rightarrow \, ^2\!S_{1/2}$ transition from~\eref{W} yielding
$|\langle ^2\!S_{1/2} ||d|| ^2\!P^o_{3/2} \rangle| \approx 3.36(3)$
a.u. The same experimental value for this reduced ME was quoted
in~\cite{DzuFla11}.

In~\tref{E1}, we present the results obtained for the reduced MEs of
the electric dipole moment operator $d$ in the DF approximation and
list the MBPT(HO), RPA, and other corrections. We emphasize that the
core-Bruckner, SR, and normalization corrections are small in this
case, and we do not present them separately. The sum of these
corrections is given in the table in the row labeled ``Other''. We
also calculated the E1 matrix elements using the \textit{ab initio}
all-order method~\cite{SafJoh08}. These values are listed in the row
labeled ``All-order''. These results include the dominant SR,
normalization, and other corrections to all orders. The MBPT(HO) and
single-double all-order results are in close agreement.

%##############################################################################################################
\begin{table*}
\caption{The breakdown of different contributions to the reduced MEs
of the electric dipole moment operator $d$ (in a.u.). First row gives
the DF values. The 2nd and 3rd rows give the MBPT(HO) and RPA
corrections, respectively. The row labeled ``Other'' is the sum of the
core-Brueckner, structural radiation, and normalization corrections.
The row labeled ``Total'' gives the final numbers. The results of the SD
all-order calculation are given in the row labeled ``All-order''. The
values are compared with the experimental and other
theoretical
%mgk>>>
~\cite{SafSaf09,DzuFla11,SD11}
%mgk<<<
results.}

\label{E1}

\begin{ruledtabular}
\begin{tabular}{lccccc}
           &$|\langle ^2\!S_{1/2} ||d|| ^2\!P^o_{1/2} \rangle|$
                     &$|\langle ^2\!S_{1/2} ||d|| ^2\!P^o_{3/2} \rangle|$
                               &$|\langle ^2\!D_{3/2} ||d|| ^2\!P^o_{1/2} \rangle|$
                                          &$|\langle ^2\!D_{3/2} ||d|| ^2\!P^o_{3/2} \rangle|$
                                                     &$|\langle ^2\!D_{5/2} ||d|| ^2\!P^o_{3/2} \rangle|$\\
\hline
DF         &  3.24   &  4.54   &  3.86    &  1.70    &  5.20 \\
MBPT(HO)   & -0.16   & -0.24   & -0.47    & -0.23    & -0.61 \\
RPA        & -0.33   & -0.42   & -0.32    & -0.12    & -0.36  \\
Other      &  0.002  & -0.05   & -0.006   &  0.001   & -0.002 \\
Total      &  2.75   &  3.83   &  3.06    &  1.35    &  4.23 \\[0.1pc]
All-order  &  2.64   &  3.71   &  2.98    &  1.32    & \\[0.1pc]
Experiment &  2.471(3)\footnotemark[1]
                     &  3.36(3)\footnotemark[2]
                                          & 2.97(4)\footnotemark[1] & \\[0.1pc]
Reference~\cite{SafSaf09}
           & 2.68    &  3.77   & 2.97     & 1.31      & 4.12 \\
Reference~\cite{DzuFla11}
           & 2.72    &  3.84   & 3.09     & 1.36      &  \\
%mgk>>>
Reference~\cite{SD11}
           & 2.72    &  3.83   & 3.06     & 1.35      & 4.23 \\
%mgk<<<
\end{tabular}
\footnotemark[1]{References~\cite{OlmHayMat09,OlmYouMoe07}};
\footnotemark[2]{Reference~\cite{PinBerJi94}, see also explanation in the text}.
\end{ruledtabular}
\end{table*}
%####################################################################

\subsection{Parity-nonconserving amplitude}
We carried out the calculation of the spin-dependent PNC
amplitudes for the $^2\!S_{1/2} \rightarrow \, ^2\!D_{3/2,5/2}$
transitions. The Hamiltonian describing the main part of the nuclear
spin-dependent PNC electron-nuclear interaction can be written as
follows
%------------------------------------------------------------------
\begin{eqnarray}
H_{\rm SD} = \frac{G_F}{\sqrt{2}}                             %  (1)
\frac{\varkappa}{I} {\bm \alpha} {\bm I} \rho({\bm r}),
\label{e1}
\end{eqnarray}
%------------------------------------------------------------------
where $G_F \approx 2.2225 \times 10^{-14}$ a.u. is the Fermi constant of
the weak interaction, $\varkappa$ is the dimensionless coupling constant,
$\bm\alpha=\left(
\begin{array}
[c]{cc}%
0 & \bm\sigma\\
\bm\sigma & 0
\end{array}
\right)$ and $\gamma_5$ are the Dirac matrices, $\bm I$ is the
nuclear spin, and $\rho({\bf r})$ is the nuclear density distribution.

We consider the nucleus to be a uniformly charged sphere. Then,
$$\rho({\bm r}) = \frac{3}{4 \pi R^3} \, \theta (R-r).$$
The root-mean-square charge radius is
$r_{\rm rms}$ = 5.2891 fm~\cite{Ang04} and, respectively, the nuclear
radius $R = \sqrt{5/3} \ r_{\rm rms}\approx 6.828$ fm.

If $|i \rangle$ and $|f \rangle$ are the initial and final atomic
states of the same nominal parity, then taking into account the
nuclear SD part of the PNC interaction in the lowest nonvanishing
order, one  can write  the electric dipole transition matrix element
as
%------------------------------------------------------------------
\begin{eqnarray}
   \langle f | d_{q,\rm SD}  | i \rangle  &=&  \sum_{n} \left[      %(4)
\frac{\langle f | d_q | n  \rangle
      \langle n | H_{\rm SD} | i \rangle}{E_i - E_n}\right.
\nonumber \\
      &+&
\left.\frac{\langle f | H_{\rm SD} | n  \rangle
      \langle n | d_q | i \rangle}{E_f - E_n} \right],
\label{e2}
\end{eqnarray}
%------------------------------------------------------------------
where $|a \rangle \equiv |J_a F_a M_a \rangle$, ${\bm F} = {\bm I} + {\bm J}$
is the total angular momentum, $M$ is the projection of ${\bf F}$,
and $H_{\rm SD}$ is given by~\eref{e1}.

%####################################################################
\begin{table*}
\caption{The nuclear spin-dependent PNC amplitude (in units $i
\varkappa \cdot 10^{-12}\, |e| a_0$),
where $a_0$ is the Bohr radius.  The values obtained in the DF and
DF+MBPT(HO) approximations are listed in the columns labeled ``DF'' and ``MBPT(HO)''.
The RPA and other corrections are included in the results listed in
the column labeled ``RPA+other''. The rows labeled ``core'' show contributions of
the core excitations. The final values (given in the rows labeled ``Total'') are
compared with the results obtained in Ref.~\cite{DzuFla11}.}
\label{PNC}
\begin{ruledtabular}
\begin{tabular}{ccrccccccc}
$F_f$ & $F_i$ & &  DF   & MBPT(HO)& RPA+other & Ref.~\cite{DzuFla11}\\
\hline
  1   &   0   & $\langle ^2\!D_{3/2},F_f ||d \cdot R_1 \cdot H_{\rm SD}|| ^2\!S_{1/2},F_i \rangle$
                & 6.2   &   6.6   &  6.9      & \\
      &       & $\langle ^2\!D_{3/2},F_f ||H_{\rm SD} \cdot R_2 \cdot d|| ^2\!S_{1/2},F_i \rangle$
                &  0    &    0    & -5.1      &\\
      &       &  core
                & 0.7   &   0.7   &  0.8      &\\
      &       & Total: $\langle ^2\!D_{3/2},F_f || d_{\rm SD} || ^2\!S_{1/2},F_i \rangle$
                & 6.9   &   7.3   &  2.6      &  3.1(1.9)   \\[0.3cm]
%\hline
  1   &   1   & $\langle ^2\!D_{3/2},F_f ||d \cdot R_1 \cdot H_{\rm SD}|| ^2\!S_{1/2},F_i \rangle$
                & 1.5   &   1.6   &  1.5      &\\
      &       & $\langle ^2\!D_{3/2},F_f ||H_{\rm SD} \cdot R_2 \cdot d|| ^2\!S_{1/2},F_i \rangle$
                &  0    &    0    & -3.2      &\\
      &       & core
                & 0.2   &   0.2   &  0.2      &\\
      &       & Total:  $\langle ^2\!D_{3/2},F_f || d_{\rm SD} || ^2\!S_{1/2},F_i \rangle$
                & 1.7   &   1.8   & -1.5      &  -1.3(4)    \\[0.3cm]
%\hline
  2   &   1   & $\langle ^2\!D_{3/2},F_f ||d \cdot R_1 \cdot H_{\rm SD}|| ^2\!S_{1/2},F_i \rangle$
                & -3.3  &  -3.5   & -3.6      &\\
      &       & $\langle ^2\!D_{3/2},F_f ||H_{\rm SD} \cdot R_2 \cdot d|| ^2\!S_{1/2},F_i \rangle$
                &  0    &    0    &  1.8      &\\
      &       &  core
                & -0.4  &  -0.4   & -0.4      &\\
      &       & Total: $\langle ^2\!D_{3/2},F_f || d_{\rm SD} || ^2\!S_{1/2},F_i \rangle$
                & -3.7  &  -3.9   & -2.2      &  -2.6(1.3)   \\[0.3cm]
%\hline
  2   &   1   & $\langle ^2\!D_{5/2},F_f ||d \cdot R_1 \cdot H_{\rm SD}|| ^2\!S_{1/2},F_i \rangle$
                &       &         &  0.2      &\\
      &       & $\langle ^2\!D_{5/2},F_f ||H_{\rm SD} \cdot R_2 \cdot d|| ^2\!S_{1/2},F_i \rangle$
                &       &         & -1.1      &\\
      &       &  core
                &       &         & -0.1      &\\
      &       & Total: $\langle ^2\!D_{5/2},F_f || d_{\rm SD} || ^2\!S_{1/2},F_i \rangle$
                &       &         & -1.0      &      %\\[0.3cm]
\end{tabular}
\end{ruledtabular}
\end{table*}
%####################################################################
The expression for the reduced ME of $d_{\rm SD}$ was derived
in~\cite{PorKoz01} and is given by
%------------------------------------------------------------------
\begin{eqnarray}
     \langle J_f &F_f& || d_{\rm SD} || J_i F_i \rangle \nonumber \\
     &=&  \sqrt{I(I+1)(2I+1)(2F_i+1)(2F_f+1)} \nonumber \\
    &\times&
     \sum_{n} \left[ (-1)^{J_f - J_i}
     \left\{ \begin{array}{ccc}
     J_n  &  J_i  &   1    \\
      I   &   I   &  F_i   \\                                    % (7)
     \end{array} \right\}
     \left\{ \begin{array}{ccc}
      J_n  &  J_f  &  1   \\
      F_f  &  F_i  &  I   \\
     \end{array} \right\} \right. \nonumber \\
  &\times& \frac{ \langle J_f || d || n, J_n \rangle
     \langle n, J_n || H_{\rm SD} || J_i \rangle }{E_n - E_i} \nonumber \\
  &+&
     (-1)^{F_f - F_i}
     \left\{ \begin{array}{ccc}
     J_n  &  J_f  &   1    \\
      I   &   I   &  F_f   \\
     \end{array} \right\}
     \left\{ \begin{array}{ccc}
     J_n  &  J_i  &  1   \\
     F_i  &  F_f  &  I   \\
     \end{array} \right\} \nonumber \\
 &\times&
     \left. \frac{\langle J_f || H_{\rm SD} ||n,J_n \rangle
            \langle n,J_n || d ||J_i \rangle}{E_n - E_f}  \right].
\label{Eq:dsd}
\end{eqnarray}
%--------------------------------------------------------------------------------------------------
For the $^2\!S_{1/2} \rightarrow \, ^2\!D_J$ transitions, where $J$=
3/2 and 5/2, in $^{171}$Yb ($I=1/2$) we obtain from \eref{Eq:dsd}
%--------------------------------------------------------------------------------------------------
\begin{eqnarray}
\label{e5}
     \langle ^2\!D_J,&F_f& || d_{\rm SD} || ^2\!S_{1/2},F_i \rangle \nonumber \\
     &=& \sqrt{\frac{3\,(2F_i+1)(2F_f+1)}{2}} \nonumber \\
    &\times&
     \sum_{n} \left[
     (-1)^{J-1/2} \left\{ \begin{array}{ccc}
     J_n  &  1/2  &   1    \\
     1/2  &  1/2  &  F_i   \\                                    % (7)
     \end{array} \right\}
     \left\{ \begin{array}{ccc}
      J_n  &  J    &  1   \\
      F_f  &  F_i  &  I   \\
     \end{array} \right\} \right. \nonumber \\
  &\times& \frac{ \langle ^2\!D_J || d || n,J_n \rangle
     \langle n,J_n || H_{\rm SD} || ^2\!S_{1/2} \rangle }{E_n - E_{^2\!S_{1/2}}} \nonumber \\
  &+&
     (-1)^{F_f - F_i}
     \left\{ \begin{array}{ccc}
     J_n  &  J      &   1    \\
     1/2  &  1/2  &  F_f   \\
     \end{array} \right\}
     \left\{ \begin{array}{ccc}
     J_n  & 1/2  &  1   \\
     F_i  &  F_f  &  I   \\
     \end{array} \right\} \nonumber \\
 &\times&
     \left. \frac{\langle ^2\!D_J || H_{\rm SD} ||n,J_n \rangle
            \langle n,J_n || d || ^2\!S_{1/2} \rangle}{E_n - E_{^2\!D_J}}  \right].
\label{6s5d}
\end{eqnarray}
%--------------------------------------------------------------------------------------------------
For subsequent calculations it is convenient to write
%--------------------------------------------------------------------------------------------------
\begin{eqnarray}
     \langle ^2\!D_J,&F_f& || d_{\rm SD} || ^2\!S_{1/2},F_i \rangle \nonumber \\
     &=& \langle ^2\!D_J,F_f || d \cdot R_1 \cdot H_{\rm SD} || ^2\!S_{1/2},F_i \rangle \nonumber \\
      &+&  \langle ^2\!D_J,F_f || H_{\rm SD} \cdot R_2 \cdot  d || ^2\!S_{1/2},F_i \rangle,
\label{6s5d_2}
\end{eqnarray}
%--------------------------------------------------------------------------------------------------
where we denote  the terms involving summations over $n$ by $R_1$ and
$R_2$.

To calculate the nuclear spin-dependent PNC amplitude defined by
Eq.~(\ref{6s5d}), one needs to sum over all possible intermediate
states or to solve the corresponding inhomogeneous equation. Here, we
solve the inhomogeneous equation using the Sternheimer-Dalgarno-Lewis
method~\cite{Ste50,DalLew55} in the valence sector. The results
obtained in different approximations are presented in \tref{PNC}.

We carried out the calculations in the DF and DF+MBPT(HO) (i.e., including
the higher orders of the MBPT) approximations. Note that in these
approximations $\langle ^2\!D_J ||H_{\rm SD}|| ^2\!P^o_{J'} \rangle = 0$
and, respectively, the second term in~\eref{6s5d_2} is also zero.

Then, we solved the RPA equations which is equivalent to the
summation of the corresponding many-body diagrams to all orders for
both $d$ and $H_{\rm SD}$ operators in~\eref{6s5d}. Smaller
contributions that include core-Brueckner, structural radiation, and
normalization corrections were also taken into account. When the RPA
corrections are included, the intermediate $nP^o_{3/2}$ states also
contribute to the spin-dependent PNC amplitude drastically increasing
(in absolute value) the second term in~\eref{6s5d_2}.

The initial and final states are the many-electron states. Therefore,
we need to account for the core excitations. This contribution (labeled
as ``core'' in~\tref{PNC}) was calculated in the DF and RPA approximations.

\tref{PNC} (see the column labeled ``RPA+other'')
illustrates that the two terms in~\eref{6s5d_2} are
comparable in their magnitude but have the opposite sign for all $F_i
\rightarrow F_f$ transitions. Therefore, they partially cancel
each other. Unfortunately, the accuracy of the calculation of the
second term is expected to be rather poor since the intermediate
$^2\!P^o_{3/2}$ state contributes to the second term at the level of
90\%. The quality of the wave function for this state near the
nucleus is expected to be low because of the discrepancies of the
theoretical and experimental values for the magnetic dipole hfs constant
$A(^2\!P^o_{3/2})$ discussed above. Because of the significant
cancellation between terms, the final numbers in~\tref{PNC} are
expected to only give an order of magnitude estimates of the
spin-dependent PNC amplitudes.

A similar single-electron approach was used by Dzuba and Flambaum
in~\cite{DzuFla11} for calculating the PNC amplitude. They have rescaled
the {\it ab initio} value of the ME $\langle n |H_{\rm SD}| m \rangle$ as
\begin{equation}
\langle n| H_{\rm SD}| m \rangle_{\rm rescaled} =
\sqrt{\frac{A_{\rm exp}(n)\,A_{\rm exp}(m)} {A_{\rm th}(n)\,A_{\rm th}(m)}}
\langle n| H_{\rm SD}| m \rangle,
\end{equation}
where $A_{\rm exp}(k)$ and $A_{\rm th}(k)$ are the experimental and
theoretical values of the magnetic dipole hfs constant of the state
$k$. The assumption that
\begin{equation}
\langle ^2\!D_{3/2}| H_{\rm SD}| ^2\!P^o_{3/2} \rangle_{\rm rescaled}
\sim \sqrt{A_{\rm exp}(^2\!D_{3/2})\,A_{\rm exp}(^2\!P^o_{3/2})}
\end{equation}
may not hold for the Yb$^+$ ions due to the mixing of the $4f^{14}6p
\,\, ^2\!P^o_{3/2}$ and $4f^{13}5d6s \,\, ^3\![3/2]^o_{3/2}$ states.
An admixture of the configuration $4f^{13}5d6s$ to the leading
configuration $4f^{14}6p$ of the $^2\!P^o_{3/2}$ state leads to an
additional contribution to the hfs constant $A(^2\!P^o_{3/2})$ which
is proportional to $\langle 4f^{13}5d6s |H_{\rm hfs}| 4f^{13}5d6s
\rangle$. This is a large contribution. However,  the configuration
$4f^{13}5d6s$ does not contribute {\it explicitly} to $\langle
4f^{14}5d |H_{\rm SD}| 4f^{13}5d6s \rangle$ because the one-electron
ME $\langle  4f |H_{\rm SD}| 6s \rangle = 0$.
Our values are in agreement with the  results of Dzuba and Flambaum~\cite{DzuFla11}
within the estimated uncertainties.
%==============================================
\section{15-electron configuration interaction}
\label{CI}
%==============================================
 We demonstrated  in the preceding sections that the single-electron method
sometimes fails to correctly predict certain properties of the Yb$^+$
ions due to mixing of configurations outside of the monovalent states
space. This mixing can be taken into account  within the framework of
the 15-electron CI. In this approach, the $4f$ electrons are also
considered as the valence electrons.

We again start from the solution of the Dirac-Fock equations, but the
construction of the DF orbitals is more complicated than in the
monovalent approximation described in the preceding section. The
odd-parity low-lying levels belong to three different configurations,
$4f^{13} 6s^2$, $4f^{14} 6p$, and $4f^{13} 5d6s$. Therefore, if we
construct the basis set in a standard way, i.e., in the V$^{N-1}$
approximation,  the $4f^{13} 6s^2$ and $4f^{13} 5d6s$ states will
have much higher energy than the $4f^{14} 6p$ states and,
respectively, there will be no mixing interaction between these
configurations. To avoid this problem, we carry out the initial
self-consistency procedure for the [1$s^2$,...,4$f^{14}$, 6$p$]
configuration. Then, all electrons were frozen and two electrons (one
from the $4f$ shell and another one from the $6p$ shell) were moved
to the $6s$ shell. Thus, the $6s$ orbital was constructed for the
$4f^{13} 6s^2$ configuration. Next, all electrons were frozen again
and one electron from the $6s$ shell was moved to the $5d$ shell. The
$5d_{3/2,5/2}$ orbitals were constructed for the $4f^{13} 5d6s$
configuration.

The basis set used in the CI calculations included virtual orbitals
up to $8s$, $8p$, $7d$, $7f$, and $5g$. The virtual orbitals were
constructed as described in~\cite{Bog91,KozPorFla96}. As a result,
the basis set we used for these calculations is rather short since
the size of the configuration space grows very rapidly with the
increase of the basis set. The configuration space was formed by
allowing single and double excitations for the even-parity states
from the configurations $4f^{14} 6s$ and $4f^{14} 5d$ and for the odd-parity
states from the configurations $4f^{14} 6p$, $4f^{13} 6s^2$, and
$4f^{13} 5d6s$. To check convergence of the CI, we calculated the
low-lying energy levels for three cases: 1) including the
single and double excitations to the shells $6s$, $6p$, $5d$, and
$5f$ (we designate it [$6sp5df$]), 2) including the single and double
excitations to [$7sp6df5g$], and 3) including the single and double
excitations to [$8sp7df5g$]. In the last case the configuration
space consisted of $\sim 2\,300\,000$ determinants and calculation of
the energy levels was rather lengthy.

\subsection{Energy levels}
The low-lying energy levels were calculated using the three CI spaces described above.
The results are presented in~\tref{tab_15E}.
%#######################################################################################
\begin{table*}
\caption{The energy levels of the low-lying excited states counted
from the ground state (in cm$^{-1}$). The columns [$6sp5df$],
[$7sp6df5g$], and [$8sp7df5g$] give results obtained using different
sets of the configurations described in the text. The results
of the single-double all-order calculations
are presented in the column labeled ``All-order'' for comparison. The experimental
energy levels~\cite{NIST} are presented in the column labeled ``Experiment''.}
\label{tab_15E}
\begin{ruledtabular}
\begin{tabular}{llcccccc}
 Config.      &     Term            &[$6sp5df$]
                                              &[$7sp6df5g$]
                                                         &[$8sp7df5g$]
                                                                    & All-order & Experiment \\
\hline \\
$4f^{14}6s$   & $^2\!S_{1/2}$       &    0    &    0     &    0     &    0     &    0  \\
$4f^{14}5d$   & $^2\!D_{3/2}$       &  29978  &  24237   &  24615   &  22820   &  22961 \\
              & $^2\!D_{5/2}$       &  30283  &  25068   &  25464   &  24261   &  24333 \\
[0.2cm]
$4f^{13}6s^2$ & $^2\!F^o_{7/2}$     &  24621  &  26735   &  26760   &          &  21419 \\
$4f^{13}5d6s$ & $^3[3/2]^o_{5/2}$   &  22977  &  25992   &  26201   &          &  26759 \\
$4f^{14}6p$   & $^2\!P^o_{1/2}$     &  21266  &  24057   &  24289   &  27945   &  27062 \\
$4f^{13}5d6s$ & $^3[3/2]^o_{3/2}$   &  26232  &  28782   &  28973   &          &  28758 \\
              & $^3[9/2]^o_{9/2}$   &  27595  &  30169   &  30364   &          &  30224 \\
$4f^{14}6p$   & $^2\!P^o_{3/2}$     &  24288  &  27093   &  27324   &  31481   &  30392 \\
$4f^{13}5d6s$ & $^3[11/2]^o_{11/2}$ &  27732  &  30412   &  30616   &          &  30563 \\
              & $^3[11/2]^o_{13/2}$ &  28297  &  31165   &  31407   &          &  31632 \\
              & $^3[5/2]^o_{7/2}$   &  29928  &  32329   &  32531   &          &  31979 \\
              & $^3[5/2]^o_{5/2}$   &  30323  &  32730   &  32939   &          &  32731 \\
\end{tabular}
\end{ruledtabular}
\end{table*}
%####################################################################
%#######################################################################################
\begin{table*}
\caption{The magnetic dipole hfs constants $A$ (in MHz). The columns
[$6sp5df$], [$7sp6df5g$], and [$8sp7df5g$] give results obtained
using different sets of the configurations described in the text.
The MBPT(HO) results are presented in the column labeled ``MBPT(HO)''
for comparison. The available  experimental values are given in column labeled
``Experiment''.}
\label{15HFS}
\begin{ruledtabular}
\begin{tabular}{llcccccr}
Configuration &     Term            &[$6sp5df$]
                                              &[$7sp6df5g$]
                                                         &[$8sp7df5g$]
                                                                    & MBPT(HO) & Experiment \\
\hline \\
$4f^{14}6s$   & $^2\!S_{1/2}$       &  18430  &  12916   &  12679   &  13091   & 12645(2)\footnotemark[1]\\
$4f^{14}5d$   & $^2\!D_{3/2}$       &    690  &    425   &    455   &    489   &   430(43)\footnotemark[2] \\
              & $^2\!D_{5/2}$       &    252  &    154   &    164   &    -96   &  -63.6(7)\footnotemark[3] \\
[0.2cm]
$4f^{13}6s^2$ & $^2\!F^o_{7/2}$     &    946  &    973   &    977   &          & 905.0(5)\footnotemark[4] \\
$4f^{13}5d6s$ & $^3[3/2]^o_{5/2}$   &   4520  &   3848   &   3841   &          &       \\
$4f^{14}6p$   & $^2\!P^o_{1/2}$     &   1264  &   1437   &   1532   &   2371   & 2104.9(1.3)\footnotemark[1] \\
$4f^{13}5d6s$ & $^3[3/2]^o_{3/2}$   &   -964  &   -742   &   -798   &          &       \\
              & $^3[9/2]^o_{9/2}$   &   -719  &   -436   &   -430   &          &       \\
$4f^{14}6p$   & $^2\!P^o_{3/2}$     &    783  &    763   &    765   &    330   & 877(20)\footnotemark[5] \\
$4f^{13}5d6s$ & $^3[11/2]^o_{11/2}$ &   1427  &   1347   &   1365   &          &       \\
              & $^3[11/2]^o_{13/2}$ &   2036  &   1782   &   1776   &          &       \\
              & $^3[5/2]^o_{7/2}$   &   3208  &   2776   &   2770   &          &       \\
              & $^3[5/2]^o_{5/2}$   &   1518  &   1246   &   1237   &          &       \\
\end{tabular}
\footnotemark[1]{Reference~\cite{MarGouHan94}};
\footnotemark[2]{Reference~\cite{Ita06} and references therein};
\footnotemark[3]{Reference~\cite{RobTayGat99}};
\footnotemark[4]{Reference~\cite{TayRobMac99}};
\footnotemark[5]{Reference~\cite{BerMal92}}.
\end{ruledtabular}
\end{table*}
%####################################################################

We were able to reproduce  the low-lying even- and odd-parity states
belonging to {\it five} different configurations $4f^{14}6s$,
$4f^{14}5d$, $4f^{14}6p$, $4f^{13}6s^2$, and $4f^{13}5d6s$ (the
column [$8sp7df5g$] in Table~\ref{tab_15E}) reasonably well.
%##############################################################################################################
\begin{table}
\caption{The absolute values of the reduced MEs $|\langle \gamma'
||d|| \gamma \rangle|$ (in a.u.), where $\gamma$ are the even-parity
states and $\gamma'$ are the odd-parity states.}

\label{15E1}

\begin{ruledtabular}
\begin{tabular}{lccc}
                                     &$f^{14}6s \,\, ^2\!S_{1/2}$
                                               &$f^{14}5d \,\, ^2\!D_{3/2}$
                                                          &$f^{14}5d \,\, ^2\!D_{5/2}$ \\
\hline
$f^{14}6p   \,\, ^2\!P^o_{1/2}$      &  2.51   &  2.53   &        \\
$f^{14}6p   \,\, ^2\!P^o_{3/2}$      &  3.32   &  1.05   &  3.27  \\
$f^{13}6s^2 \,\, ^2\!F^o_{7/2}$      &         &         &  0.063  \\
$f^{13}5d6s \,\, ^3\![3/2]^o_{5/2}$  &         & 0.00075 &  0.0037 \\
$f^{13}5d6s \,\, ^3\![3/2]^o_{3/2}$  &  1.10   &  0.27   &  0.86  \\
\end{tabular}
\end{ruledtabular}
\end{table}
%####################################################################
The theoretical energy levels for the $4f^{14}6p \,\, ^2\!P^o_J$
states are located deeper than the experimental levels. It is not
surprising since the initial self-consistency Dirac-Fock procedure
was carried out for this configuration. The levels of the
$4f^{13}5d6s$ configuration are in reasonable agreement with the
experimental data. The $4f^{14}5d \,\, ^2\!D_J$ states are lying
5-7\% higher than the experimental levels. This is also expected
because, as we mentioned above, the $5d$ orbital was constructed not
for the $4f^{14}5d$ configuration but for the $4f^{13} 5d6s$
configuration. We observe the worst agreement with the experiment for
the $4f^{13}6s^2 \,\, ^2\!F^o_{7/2}$ state. A reason is a particular
sensitivity of this state to the configuration interaction. It was
confirmed by calculations carried out with other (smaller) sets of
configurations (not included in~\tref{15E1}). We assume that more
configurations have to be taken into account to reproduce this
energy level with good precision.

It is worth noting that it was essential to include the $5g$ shell
into consideration as illustrated by the comparison of the columns
[$6sp5df$] and [$7sp6df5g$] in~\tref{tab_15E}. Most of the observed
differences in the energy levels listed in these two columns are due
to including configurations involving the $5g$ orbitals into the CI
space. A number of levels are very sensitive to these configurations.
An addition of the $8s$, $8p$, $7d$, and $7f$ shells (compare the
columns [$7sp6df5g$] and [$8sp7df5g$] in the table) led  to much
smaller changes in the energy levels. Comparison of these three sets
appears to indicate that further extension of the CI space (which
will be extremely time-consuming)  will not lead to any qualitative
changes for a majority of the states.
The results obtained in the framework of the single-double all-order
approach are presented in~\tref{15E1} for comparison.
%####################################################################
\begin{table*}
\caption{Comparison of the reduced MEs of the electric dipole moment
found in the single-electron approach (the row DF+MBPT) and in the
15-electron CI approach (the row 15-el. CI). The values in the row
DF+MBPT include the MBPT(HO), RPA and smaller corrections. The
experimental values are presented in third row.}

\label{E1_comp}

\begin{ruledtabular}
\begin{tabular}{lccccc}
           &$|\langle ^2\!S_{1/2} ||d|| ^2\!P^o_{1/2} \rangle|$
                     &$|\langle ^2\!S_{1/2} ||d|| ^2\!P^o_{3/2} \rangle|$
                               &$|\langle ^2\!D_{3/2} ||d|| ^2\!P^o_{1/2} \rangle|$
                                          &$|\langle ^2\!D_{3/2} ||d|| ^2\!P^o_{3/2} \rangle|$
                                                     &$|\langle ^2\!D_{5/2} ||d|| ^2\!P^o_{3/2} \rangle|$\\
\hline
DF+MBPT    &  2.75   &  3.83   &  3.06    &  1.35    &  4.23 \\
15-el. CI  &  2.51   &  3.32   &  2.53    &  1.05    &  3.27 \\[0.1pc]

Experiment &  2.471(3)\footnotemark[1]
                     &  3.36(3)\footnotemark[2]
                                          & 2.97(4)\footnotemark[1] & \\[0.1pc]
\end{tabular}
\footnotemark[1]{References~\cite{OlmHayMat09,OlmYouMoe07}};
\footnotemark[2]{Reference~\cite{PinBerJi94}, see also explanation in
the text}.
\end{ruledtabular}
\end{table*}
%###################################################################

In Sec.~\ref{single}, we discussed a poor agreement between the
experimental value of the magnetic dipole hfs constant $A(4f^{14}6p
\,\,^2\!P^o_{3/2})$ and the value obtained in the single-electron
approach. A strong interaction of this state with the closely lying
$4f^{13}5d6s \,\, ^3[3/2]^o_{3/2}$ state was suggested  as a possible
reason of this disagreement. Our calculation with the [$8sp7df5g$] CI
space reproduced the difference between the energies of the
$^2\!P^o_{3/2}$ and $^3[3/2]^o_{3/2}$ states almost perfectly
(1649~cm$^{-1}$), though the order of the levels was not correct. The
experimental difference is 1634~cm$^{-1}$. It makes us confident that
the configuration mixing of these two states is taken into account
sufficiently correctly. We would like to stress that our calculations
are purely {\it ab initio}. No semi-empirical parameters were used in
the framework of this approach.

In the next subsection, we present the values of the magnetic dipole
hfs constants and E1 transition amplitudes between the low-lying
states. We compare these results with those obtained in the
single-electron approximation and discuss the role of the mixing of
monovalent and one-hole-two-particle configurations.

\subsection{hfs constants, E1 transition amplitudes, and other observables}

The values of the magnetic dipole hfs constants obtained using three
sets of configurations are listed in~\tref{15HFS}. We compare  these
results with the values listed in~\tref{HFS} obtained using the
single-electron method. We will discuss the results obtained for the
largest [$8sp7df5g$] CI space.

For the even-parity states, the hfs constants found in the
15-electron CI are close to the values obtained at the MBPT(HO) stage
(see~\tref{HFS}). Such an agreement looks reasonable. The CI results
include the correlation corrections between the $4f$ and other
valence electrons. In the single-electron approach these
are core-valence correlations. At the same time the core excitations
from all shells up to $4f$ are completely disregarded in the
15-electron CI approach.  The configuration mixing does not play very
significant role for the even-parity states considered here because
the even states with unfilled $4f$ shell are located rather far from
the ground- and  $^2\!D_{3/2,5/2}$ states.

The value of $A(^2\!P^o_{1/2})$ is very close to the result obtained
in the single-electron approach in the DF approximation. Our analysis
shows that in the many-electron case the contributions of all
electrons (except the $6p_{1/2}$) nearly cancels with each other and
the final value is determined almost completely by the contribution
of the $6p_{1/2}$ electron.

The most significant disagreement between the experimental and
theoretical hfs constants in the single-electron approach (by a
factor of 2.7) was found for the $^2\!P^o_{3/2}$ state. This problem
is resolved in the many-electron calculation.  The 15-electron CI
gives the value 765 MHz which differs from the experiment by only
13\% (see \tref{15HFS}, column [$8sp7df5g$]). The admixture of the
$4f^{13}5d6s \,\, ^3\![3/2]^o_{3/2}$ state to the $4f^{14}6p \,\,
^2\!P^o_{3/2}$ state leads to an appearance of the contribution from
the one-electron ME $\langle 6s |H_{\rm hfs}| 6s \rangle$ which
strongly affects the value of this constant. Based on the results
obtained in the single-electron approach (see~\tref{HFS}, the row
RPA), we estimate that the core-valence correlation corrections will
increase this number making it even closer to the experimental
result. Thus, if the interaction between the $^2\!P^o_{3/2}$ and
$^3[3/2]^o_{3/2}$ states is taken into account, the value of
$A(^2\!P^o_{3/2})$ turns out to be in a good agreement with the
experimental result.

We found only one experimental result for the states with the
unfilled $4f$ shell listed in~\tref{15HFS} (for the $4f^{13}6s^2 \,\,
^2\!F^o_{7/2}$ state). Our value agrees with the experiment at the
level of 7\%. An assignment of uncertainties to the hfs constants of
these states is not trivial. One source of errors is the
insufficiently large CI space. The corresponding uncertainties may be
estimated by a comparison of the results obtained for the
[$7sp6df5g$] and [$8sp7df5g$] CI spaces. We see that the difference
is not so large (at the level of a few percent). Another source of
uncertainties  is the core-valence correlations omitted in this
approach. A magnitude of these corrections can be estimated using the
results obtained in DF+MBPT method (see~\tref{HFS}). For the large
hfs constants $A(^2\!S_{1/2})$ and $A(^2\!P^o_{1/2})$, they
contribute $\sim$ 30-35\%. In the 15-electron approach these
corrections are expected to be smaller because the CI core
[$1s^2$,...,$5p^6$] contains less electrons and is more ``hard'' than
the  [$1s^2$,...,$5p^6$, $4f^{14}$] core. Thus, we estimate accuracy
of our results to be at the level of 25-30\%. For comparison
we also present in~\tref{15HFS} the results obtained
in the framework of the single-electron MBPT(HO) approach.

We also calculated the E1 transition amplitudes for the low-lying
states. The reduced MEs of the electric dipole operator $d$ are given
in~\tref{15E1}.  It is instructive to compare the results obtained by
the single-electron and 15-electron CI methods. This comparison is
carried out in~\tref{E1_comp}.

The results indicate the following trend. The values of the MEs of
the transitions between the ground and $^2\!P^o_J$ states are closer
to the experimental values in the 15-electron CI approach, while
DF+MBPT method gives better agreement with the experiment for the
$^2\!P^o_{1/2} - \, ^2\!D_{3/2}$ transition. As we noted above,
taking into account the configuration mixing is important for the
$^2\!P^o_{3/2}$ state. This mixing also manifests itself for the
reduced ME $|\langle ^2\!S_{1/2} ||d|| ^2\!P^o_{3/2} \rangle|$ though
its influence is weaker than for the hfs constant $A(^2\!P^o_{3/2})$.

We conclude that {\it all}  $|\langle ^2\!D_J ||d|| ^2\!P^o_{J'}
\rangle|$ matrix elements are obtained to better accuracy in the
single-electron method. In the single-electron approach, the $5d$
orbital was constructed for the $f^{14}5d$ configuration which is
``native'' for the $^2\!D_J$ states. In the 15-electron CI approach,
it was constructed for the $f^{13}5d6s$ configuration. Most likely,
the set of configurations used to form the wave function of the
$^2\!D_J$ states is not sufficiently large (even for the biggest CI
space that we have considered) to correctly reproduce their
properties.

In the recent work of Huntemann {\it et~al.} \cite{HunOkhLip12}, the
quadrupole moment $\Theta$ of the $4f^{13}6s^2 \,\, ^2\!F^o_{7/2}$
state was measured to be $\Theta=-0.041(5)\,e a_0^2$. The quadrupole
moment $\Theta$ of a state $|\gamma J \rangle$ (where $\gamma$
designates all quantum numbers except $J$) is determined as
\begin{equation}
\Theta = 2 \sqrt {\frac{J(2J-1)}{(2J+3)(2J+1)(J+1)}}
\langle \gamma J ||Q_2|| \gamma J \rangle,
\end{equation}
where $Q_2$ is the electric quadrupole operator.

We carried out the calculation of $\Theta(4f^{13}6s^2 \,\, ^2\!F^o_{7/2})$
for three increasing CI spaces [6sp5df], [7sp6df5g], and [8sp7df5g].
The results are presented in~\tref{tab_Q}.
%####################################################################
\begin{table}
\caption{The quadrupole moment $\Theta$ of the $4f^{13}6s^2 \,\,
^2\!F^o_{7/2}$ state (in $e a_0^2$). The results are presented for
three CI spaces [6sp5df], [7sp6df5g], and [8sp7df5g] and compared
with the experimental and another theoretical value.}
\label{tab_Q}
\begin{ruledtabular}
\begin{tabular}{ll}
                                & $\Theta$ \\
\hline
$[6sp5df]$                      & -0.40 \\
$[7sp6df5g]$                    & -0.06 \\
$[8sp7df5g]$                    & -0.20 \\
Other theory~\cite{BlyWebHos03} & -0.22 \\
Experiment~\cite{HunOkhLip12}   & -0.041(5)
\end{tabular}
\end{ruledtabular}
\end{table}
%####################################################################
As illustrated by the table, our result obtained for the largest CI
space coincide with the theoretical value of Ref.~\cite{BlyWebHos03}
and is 5 times greater (in absolute value) than the experimental result.
At the same time, we see that $\Theta$ is very sensitive to the configuration
interaction. The quadrupole moment is rather small due to large
cancelations of one-electron contributions which is expected to make
its accurate calculation more difficult. All this makes the result
obtained even for the biggest CI space $[8sp7df5g]$ rather
inconclusive. Based on our calculations, we can only roughly estimate
this quantity as $\Theta(4f^{13}6s^2 \,\, ^2\!F^o_{7/2}) \sim -0.1
\, e a_0^2$.

Finally, we note that an attempt to take into account core-valence
correlations by combining the 15-electron CI with the MBPT was
unsuccessful. The main problem which was repeatedly discussed earlier
(see, e.g.,~\cite{PorKosTup07}) is instability of the MBPT for the
mean-field potential $V^N$, which includes a large number of valence
electrons. An accounting for the MBPT corrections leads to an
appearance of huge contribution from the subtraction
diagrams~\cite{DzuFlaKoz96}. These diagrams are calculated only in
the second order of the MBPT. This is insufficient for accurate
treatment of the core-valence correlations.

This problem does not allow us to calculate the SD PNC amplitude more
accurately by the 15-electron CI method than it was done in the
single-electron approach because the matrix element  of the SD PNC
Hamiltonian $\langle ^2\!D_J || H_{\rm SD}|| ^2\!P^o_{3/2} \rangle$
is greatly increased when we include the RPA and other corrections.
To perform similar calculation in the framework of the 15-electron
CI, we need to include the subtraction diagrams into consideration,
what makes this approach very unstable.

Formulating CI+all-order approach
%~\cite{SafKozJoh09}
that can treat two-particle-one-hole states on the same footing as
the monovalent states appears to be a promising way for a development
of methodologies capable to further improve the calculation accuracy
of the Yb$^+$ properties.
%####################################################################
\section{Conclusion}
\label{Concl} To conclude, we calculated the energies,
magnetic dipole hfs constants, E1 transition amplitudes between the
low-lying states, and the nuclear spin-dependent parity-nonconserving
amplitudes for the $^2\!S_{1/2} - \, ^2\!D_{3/2,5/2}$ transitions.
Our calculations were carried out in the framework of the
single-electron DF+MBPT  method and by the 15-electron CI method.
All-order calculations were also carried out for selected properties
using the linearized single-double coupled-cluster method.

The specific character of Yb$^+$ ion manifests itself due to the
presence of the low-lying states with unfilled
$4f$ shell. A configuration interaction between them and the states
with filled $4f$ shell significantly affects the properties of both
types of states. We demonstrated this configuration mixing by
analyzing the properties of the $4f^{14}6p \,\, ^2\!P^o_{3/2}$ state.
In particular, we found that an admixture of the nearby
$4f^{13}5d6s \,\, ^3\![3/2]^o_{3/2}$ state should be taken into account.

Various contributions to the spin-dependent parity-violating amplitude
are discussed and a method to improve accuracy further is proposed.

\section*{Acknowledgments}
%####################################################################
We would like to thank V.~Dzuba and V.~Flambaum for helpful
discussions. S.G.P. is also grateful to University of New South Wales
(Australia), where this work has been started, for hospitality. This
work was supported in part by US NSF Grant No.\ PHY-1068699. The work
of M.G.K. and S.G.P. was supported in part by RFBR Grant No.\ 11-02-00943.

%#####################################################
%\bibliography{./YbII}
%\end{document}
%#####################################################
%#####################################################

\end{document}